\documentclass[aps,pra,showpacs,twocolumn,floatfix,groupedaddress]{revtex4}

\usepackage{amssymb}
\usepackage{amsmath}
\usepackage{amsfonts}
\usepackage{dcolumn,graphicx}
\usepackage{epsfig}

\begin{document}

\title{ AC Stark shift of the Cs microwave atomic clock transitions }

\author{P. Rosenbusch}
\affiliation {LNE-SYRTE, Observatoire de Paris, 75014 Paris, France}

\author{S. Ghezali}
\affiliation {Department of Physics, University Saâd Dahlab, Blida,
 and Laboratory of Quantum Electronics, University of Sciences and Technology HOUARI
 BOUMEDIENE, Bab-Ezzouar, Alger, Algeria}

\author{V. A. Dzuba}
\affiliation {
School of Physics, University of New South Wales, Sydney,
2052, Australia}

\author{V. V. Flambaum}
\affiliation{
School of Physics, University of New South Wales, Sydney,
2052, Australia}

\author{K. Beloy}
\affiliation {
Department of Physics, University of Nevada, Reno, Nevada 89557}

\author{A. Derevianko}
\affiliation {
Department of Physics, University of Nevada, Reno, Nevada 89557}

\date{\today}

\begin{abstract}


We analyze the AC Stark shift of the Cs microwave atomic clock
transition theoretically and experimentally. Theoretical and
experimental data are in a good agreement with each other.  Results
indicate the absence of a magic wavelength at which there would be
no differential shift of the clock states having zero projections of
the total angular momentum.
\end{abstract}



\maketitle

\section{Introduction}
The present definition of the unit of time, the second, is based on
the frequency of the microwave transition between two hyperfine
levels of the Cs atom. Recently,  it has been realized that the
accuracy and stability of atomic clocks can be substantially
improved by trapping atoms in optical lattices operated at a certain
``magic'' wavelength~\cite{KatTakPal03,YeKimKat08}. At this magic
wavelength, both clock levels experience the same AC Stark shift;
the clock frequency becomes essentially independent on trapping
laser intensity.

This effect was
demonstrated~\cite{TakHonHig05,LeTBaiFou06,LudZelCam08etal} for
optical clocks using strontium atoms. An extension of this idea to
microwave clocks with alkali-metal atoms Rb and Cs was considered in
Ref.~\cite{ZhoCheChe05}. A multitude of magic wavelengths for the
hyperfine transition was identified. Unfortunately, detailed
analysis presented below shows the conclusions of that paper to be
erroneous: there is no magic wavelength for Cs, at least for clock
levels with zero projections of the total angular momentum $M_F$ on
the quantizing magnetic field. In a separate
paper~\cite{FlaDzuDer08} we analyze the case of circular light
polarization and $M_F \neq 0$ levels and demonstrate that the AC
shift can be eliminated by an appropriate ``magic angle'' choice of
the direction of the magnetic field with respect to the light
propagation.

This paper presents a detailed theoretical analysis of the frequency
shift of a microwave clock involving a hyperfine transition. The
analysis requires calculation of the differential polarizability
involving third-order expressions, quadratic in the field strength
and linear in the hyperfine interaction. Evaluation of the resulting
expressions is carried out using relativistic many-body theory. The
second part of the paper reports measurements of the clock shift at
two laser wavelengths. The results of the calculations are in a good
agreement with the experimental measurements.

\section{Theory}

Here we follow the formalism of the quasi-energy states reviewed in
the context of laser-atom interaction in Ref.~\cite{ManOvsRap86}. We
start with considering the AC Stark shift $\delta E^{\left[ 2\right]
}$ in the second order of perturbation theory (quadratic in the
electric field) and then extend the formalism to the higher-order AC
Stark shift $\delta E^{\left[ 2+1\right]  }$ which takes into
account the hyperfine interaction (HFI). The latter shift appears in
the third order of perturbation theory and is quadratic in the field
amplitude and linear in the HFI. An important part of the analysis
involves the tensorial expansion of the shifts in the scalar,
vector, and tensor parts.

We are interested in transitions between two hyperfine components of
the same electronic states. Below we employ the conventional
labeling scheme for the atomic eigenstates, $\left\vert n\left(
IJ\right)  FM_{F}\right\rangle $, where $I$ is the nuclear spin, $J$
is the electronic angular momentum, and $F$ is the total angular
momentum, $\mathbf{F}=\mathbf{J}+\mathbf{I}$. $M_{F}$ is the
projection of $F$ on the quantization axis and $n$ encompasses the
remaining quantum numbers. Since the clock transitions involve the
same electronic state, we will also use a shorthand notation
$|F,M_F\rangle$. For example, the Cs fountain clock involves
transitions between hyperfine levels   $|F=4,M_F=0\rangle$ and
$|F'=3,M_F'=0\rangle$ of the $6s_{1/2}$ ground electronic state.

Under the influence of the laser each clock level is perturbed.
The clock frequency is modified by the difference in the perturbed energies
\begin{eqnarray}
\delta\nu^{\mathrm{Stark}}\left(  \omega_{L}\right)
&=&\frac{1}{h}\left[
\delta E_{n\left(  IJ\right)  F^{\prime\prime}M_{F}^{\prime\prime}%
}^{\mathrm{Stark}}\left(  \omega_{L}\right)
\right.\nonumber\\&&\left.-\delta E_{n\left( IJ\right)
F^{\prime}M_{F}^{\prime}}^{\mathrm{Stark}}\left( \omega_{L}\right)
\right] \label{Eq:ClockShift}
\end{eqnarray}
At the ``magic frequency'', this AC Stark clock shift would vanish.

\subsection{Second-order dynamic response}

This section introduces notation and reviews derivation and
tensorial analysis of the conventional second-order dynamic Stark
shift. We demonstrate that for states of total electronic angular
momentum $J=1/2$ and for $M_{F}=0$ levels (or for linear
polarization) the second-order AC Stark shift of the clock
transition vanishes.

Consider a traveling electromagnetic wave of an arbitrary polarization,
\[
\mathcal{\vec{E}}=\frac{1}{2}\mathcal{E}_{L}\hat{\varepsilon}~e^{-i\left(
\omega t-kz\right)  }+c.c.~,
\]
with the complex polarization vector parameterized by an angle $\theta$%

\begin{equation}
\hat{\varepsilon}=\mathbf{\hat{e}}_{x}\cos\theta+i\mathbf{\hat{e}}_{y}%
\sin\theta. \label{Eq:polVec}%
\end{equation}
The parametric angle $\theta$ may be related to the degrees of
linear, $l=\cos2\theta$, and circular, $A=\sin2\theta$,
polarization. Notice that the quantizing axis \thinspace$z$ is
chosen along the propagation vector $\mathbf{\hat{k}}$\textbf{ }of
the laser. The field amplitude $\mathcal{E}_{L}$ is related to the
intensity of the laser as $I_{L}=\frac
{c}{8\pi}\mathcal{E}_{L}^{2}$, or in practical units $I_{L}\left[
\frac{\mathrm{mW}}{\mathrm{cm}^{2}}\right]  \approx1.33\times\left(
\mathcal{E}_{L}\left[  \frac{\mathrm{V}}{\mathrm{cm}}\right] \right)
^{2}$.

In the dipole approximation, the coupling can be represented as ($h.c.$ is a
hermitian conjugate)%
\[
V_{E1}\left(  t\right)  \equiv-\mathcal{\vec{E}}\cdot\mathbf{D}=-\frac{1}%
{2}\mathcal{E}_{L}~\hat{\varepsilon}\cdot De^{- i\omega t}+h.c..
\]
Application of the Floquet formalism (dressed states) yields the second-order
AC shift of the atomic energy level $a$%
\begin{eqnarray*}
\delta E_{a}^{\left[  2\right]  }&=&\sum_{b}\frac{\left\vert
\langle\psi _{b}|v|\psi_{a}\rangle\right\vert ^{2}}{E_{a}-\left(
E_{b}-\omega\right)
}
+\sum_{b}\frac{\left\vert \langle\psi_{a}|v|\psi_{b}\rangle\right\vert ^{2}%
}{E_{a}-\left(  E_{b}+\omega\right)  },
\end{eqnarray*}
where
$v=-\frac{1}{2}\mathcal{E}_{L}~\hat{\varepsilon}\cdot\mathbf{D}\,$,
$\psi_{a}$ and $\psi_{b}$ being the stationary atomic states with
unperturbed energies $E_{a}$ and $E_{b}$, respectively.

Now we proceed to the conventional reduction of the polarizability
into a sum over irreducible tensor operators. Introducing the
resolvent operator
($H_{0}$ is the unperturbed atomic Hamiltonian)%
\[
R_{E_{a}}\left(  \omega\right)  =\left(  E_{a}-\hat{H}_{0}+\omega\right)
^{-1},
\]
we may recast the shift as an expectation value $\delta E_{a}^{\left(
2\right)  }=\left(  \frac{1}{2}\mathcal{E}\right)  ^{2}\langle\psi_{a}|\hat
{O}_{E1}\left(  \omega\right)  |\psi_{a}\rangle$, with%
\begin{eqnarray*}
\hat{O}_{E1}\left(  \omega\right)&=&\left( ~\hat{\varepsilon}\cdot
\mathbf{D}\right)  ^{\dagger}R_{E_{a}}\left( \omega\right) \left(
~\hat {\varepsilon}\cdot\mathbf{D}\right)\\
&&+\left( ~\hat{\varepsilon}\cdot\mathbf{D}\right) R_{E_{a}}\left(
-\omega\right)  ~\left(  \hat{\varepsilon}\cdot \mathbf{D}\right)
^{\dagger}.
\end{eqnarray*}
The order of coupling of the operators may be changed%
\begin{eqnarray*}
\hat{O}_{E1}\left(  \omega\right) &=&\sum_{K=0,1,2}\left[\left(
-1\right) ^{K}\right.\\
&&\times\left\{
\hat{\varepsilon}^{\ast}\otimes~\hat{\varepsilon}\right\}
_{K}\cdot\left\{  \mathbf{D}\otimes~R_{E_{a}}\left(  \omega\right)
\mathbf{D}\right\}  _{K}  \\&&\left.+\left\{
\hat{\varepsilon}^{\ast}\otimes~\hat{\varepsilon}\right\}
_{K}\cdot\left\{  \mathbf{D}\otimes~R_{E_{a}}\left( -\omega\right)
\mathbf{D}\right\}  _{K}\right]~,
\end{eqnarray*}
leading to the conventional decomposition into the scalar ($K=0$),
vector ($K=1$), and tensor ($K=2$) terms. Here we employed $\left\{
\hat {\varepsilon}\otimes~\hat{\varepsilon}^{\ast}\right\}
_{KM}=\left( -1\right)  ^{K}\left\{
\hat{\varepsilon}^{\ast}\otimes~\hat{\varepsilon }\right\}_{KM}$ and
the fact that $\hat{\varepsilon}$ and $\mathbf{D}$ are rank 1
tensors. The $M_{K}$ component of the compound tensor of rank $K$
composed from components of the tensors $A_{K_1}$ and $B_{K_2}$ (of
rank ${K_1}$ and ${K_2}$) is
defined as $\left\{A_{K_1}\otimes B_{K_2}~\right\}  _{KM_{K}}%
=\sum_{M_{1}M_{2}}C_{K_{1}M_{1}K_{2}M_{2}}^{KM_{K}}A_{K_{1}M_{1}}B_{K_{2}%
M_{2}}$, where $C_{{K_1}M_{1}{K_2}M_{2}}^{KM_{K}}$ are the
Clebsch-Gordan coefficients. The generalized scalar product is
defined as $\left( A_{K}\cdot B_{K}\right)  =\sum_{M_{K}}\left(
-1\right) ^{M_{K}}A_{KM_{K}}B_{K,-M_{K}}$.

Using the Wigner-Eckart theorem, a matrix element between two atomic
states
may be expressed as%
\begin{eqnarray*}
\langle FM_{F}|\hat{O}_{E1}\left(  \omega\right)
|FM_{F}^{\prime}\rangle&=&\sum_{K=0,1,2}\left(  -1\right)
^{K}\sum_{\mu}\left(  -1\right) ^{\mu }\\&&\times\left\{
\hat{\varepsilon}^{\ast}\otimes~\hat{\varepsilon}\right\}  _{K,-\mu
}   \left(  -1\right)  ^{F-M_F}\\
&&\times\left(
\begin{tabular}
[c]{lll}%
$F$ & $K$ & $F$\\
$-M_{F}$ & $\mu$ & $M_{F}^{\prime}$%
\end{tabular}
\right)  \alpha_{nF}^{\left(  K\right)  }\left(  \omega\right)  ,
\end{eqnarray*}
with the reduced polarizabilities
\begin{eqnarray}
\alpha_{nF}^{\left(  K\right)  }\left(  \omega\right)   &  =&\langle
nF||\left\{  D\otimes~R_{E_{nF}}\left(  \omega\right)  D\right\}
_{K}\nonumber\\&&+\left( -1\right)  ^{K}\left\{
D\otimes~R_{E_{nF}}\left( -\omega\right)  D\right\}
_{K}||nF\rangle\nonumber\\
&=& \sqrt{[K]}(-1)^{K+2F}\sum_{F^{\prime}}\left\{
\begin{array}
[c]{ccc}%
1 & 1 & K\\
F & F & F^{\prime}%
\end{array}
\right\} \nonumber\\&&\times\sum_{n^{\prime}}    \langle
nF||D||n^{\prime}F^{\prime}\rangle\langle
n^{^{\prime}}F^{\prime}||D||nF\rangle\nonumber \\
&&
\times\left(  \frac{1}{E_{nF}-E_{n^{^{\prime}%
}F^{^{\prime}}}+\omega}\right.\nonumber\\&&\left.+\left(  -1\right)
^{K}\frac{1}{E_{nF}-E_{n^{^{\prime
}}F^{^{\prime}}}-\omega}\right)  . \label{Eq:alphared}%
\end{eqnarray}
Here we have used the shorthand notation $[K]\equiv(2K+1)$. The
matrix element may be simplified further using specific
parametrization, Eq.~(\ref{Eq:polVec}), of the polarization vector.
Explicitly,$\left\{
\hat{\varepsilon}^{\ast}\otimes~\hat{\varepsilon}\right\}
_{00}=-\frac {1}{\sqrt{3}},$ $\left\{
\hat{\varepsilon}^{\ast}\otimes~\hat{\varepsilon }\right\}
_{1\mu}=-\frac{\sin2\theta}{\sqrt{2}}\delta_{\mu,0}~,\left\{
\hat{\varepsilon}^{\ast}\otimes~\hat{\varepsilon}\right\}
_{2\mu}=-\frac
{1}{\sqrt{6}}\delta_{\mu,0}+\frac{1}{2}\cos2\theta~\delta_{\mu,\pm2}.$

Finally, the  AC Stark energy shift reads%
\begin{eqnarray}
\delta E_{nFM_{F}}^{\left(  2\right)  }&=&-\left(  \frac{1}{2}\mathcal{E}%
_{L}\right)  ^{2}\left[  \alpha_{nF}^{s}\left(  \omega\right)
+A~\alpha _{nF}^{a}\left(  \omega\right)
\frac{M_{F}}{2F}\right.\nonumber\\&&\left.-\alpha_{nF}^{T}\left(
\omega\right) \frac{3M_{F}^{2}-F\left(  F+1\right)  }{2F\left(
2F-1\right)
}\right]  ,\label{Eq:StarkEnergyShift}%
\end{eqnarray}
with the conventional scalar, vector, and tensor polarizabilities
\begin{align}
\alpha_{nF}^{s}\left(  \omega\right)   &  =\frac{1}{\sqrt{3}}\frac{1}%
{\sqrt{[F]}}\alpha_{nF}^{\left(  0\right)  }\left(  \omega\right)
,\nonumber\\
\alpha_{nF}^{a}\left(  \omega\right)   &  =-\frac{1}{\sqrt{2}}\frac{1}%
{\sqrt{[F]}}\frac{2F}{\sqrt{F\left(  F+1\right)  }}\alpha_{nF}^{\left(
1\right)  }\left(  \omega\right)  ,\label{Eq:TraditionalContribs}\\
\alpha_{nF}^{T}\left(  \omega\right)   &  =-\frac{2}{\sqrt{6}}2F\left(
2F-1\right)  \left[  \frac{\left(  2F-2\right)  !}{\left(  2F+3\right)
!}\right]  ^{1/2}\alpha_{nF}^{\left(  2\right)  }\left(  \omega\right)
.\nonumber
\end{align}
In general, there is an off-diagonal $M_{F}=M_{F}^{\prime}\pm2$
optical coupling involving the tensor part of the polarizability. In
practice, a quantizing $\mathbf{B}$-field is applied along the
propagation of the laser wave, and as long as the off-diagonal
coupling is much smaller than the Zeeman intervals, it can be
disregarded.

Since the dipole matrix elements do not couple to the nuclear degrees of
freedom, the dependence of the reduced polarizabilities on $I$ and $F\,\ $may
be be factored out as
\[
\alpha_{nF}^{\left(  K\right)  }\left(  \omega\right)  =(-1)^{J+I+F+K}%
\;\left[  F\right]  \left\{
\begin{array}
[c]{ccc}%
J & F & I\\
F & J & K
\end{array}
\right\}  \bar{\alpha}_{nJ}^{\left(  K\right)  }\left(  \omega\right)  ,
\]
where the quantities$~\bar{\alpha}_{nJ}^{\left(  K\right)  }\left(
\omega\right)  $ are the reduced matrix elements in the $|nJM_{J}\rangle$
basis,
\begin{eqnarray*}
\bar{\alpha}_{nJ}^{\left(  K\right)  }\left(  \omega\right)
&=&\langle nJ||\left\{  D\otimes~R_{E_{nF}}\left(  \omega\right)
D\right\} _{K}\\&&+\left( -1\right)  ^{K}\left\{
D\otimes~R_{E_{nF}}\left( -\omega\right)  D\right\} _{K}||nJ\rangle.
\end{eqnarray*}
These quantities do not depend on either $I$ or $F$.

With this factorization, we can make important comments specific to
the case of $J=1/2$ (e.g., ground state of alkali-metal atoms such
as Rb and Cs). Due to the angular selection rules, the tensor
contribution (expectation value of the rank 2 tensor) vanishes and
the only contributions come from the scalar and vector parts. As the
vector part of the energy shift is proportional to $M_{F}$, for
$M_{F}=0$ clock levels only the scalar contribution remains. The
vector contribution also vanishes for the case of linear
polarization ($A=0$).

The next important step is to demonstrate that the scalar shift does not
depend on $F$. In other words, there is no clock shift at the second order.
Indeed, for the scalar term, $\alpha_{nF}^{s}\left(  \omega\right)  =\frac
{1}{\sqrt{3}}\frac{1}{\sqrt{[F]}}\alpha_{nF}^{\left(  0\right)  }\left(
\omega\right)  =\frac{1}{\sqrt{3}}\frac{1}{\sqrt{[J]}}\bar{\alpha}%
_{nJ}^{\left(  0\right)  }\left(  \omega\right)  $. This result holds for an
arbitrary $J$.

This result has a very simple explanation. As it was pointed out
above, for the case of linear polarization of laser light and
$J=1/2$ we have only a scalar contribution to the polarizability.
This contribution does not depend on the orientation of the
quantization axis (external magnetic field). Let us consider the
case when the quantization axis is directed along the laser electric
field (along the linear polarization vector ${\bf e}$). From the
symmetry of the problem it is obvious that the electron states
$|J_z=1/2>$ and $|J_z=-1/2>$ have exactly equal quadratic shifts.
The hyperfine states $|F,F_z>$ are linear combinations of these
electron states multiplied by nuclear states. Since both components,
$|J_z=1/2>$ and $|J_z=-1/2>$, of any hyperfine state have the same
shift, all hyperfine states have the same shifts, i.e. the
differential polarizability is equal to zero. To have a non-zero
differential polarizability one has to include the hyperfine
interaction. Note that the inclusion of the magnetic polarizability
does not change this conclusion. If the lattice laser frequency is
in optical range, the magnetic polarizability contribution is
suppressed by an additional factor $\mu_B^2/D^2 \sim
\alpha^2\approx(1/137)^2$ and may be neglected.

To summarize, we  arrive at the conclusion that for $J=1/2$,
$M_{F}=0$ clock levels (or for linear polarization) the second-order
AC Stark shift is zero. Since the calculations of
Ref.~\cite{ZhoCheChe05} were limited to this second-order, their
conclusions are erroneous. This is also shown in the Appendix
without using irreducible tensor algebra.


\subsection{Non-trivial effect of the hyperfine interaction}

In the previous section, the HFS interaction has served the role of
an \textquotedblleft observer\textquotedblright, as it only defined
the coupling scheme. In particular, we find that the $M_{F}=0$
levels of the  hyperfine manifold attached to the $J=1/2$ levels are
shifted identically - at that level of approximation any laser
wavelength is \textquotedblleft magic\textquotedblright, i.e., the
clock transition remains unperturbed by any laser field. The
non-trivial effect arises when we take into account the dynamic (as
opposed to the observer) role of the HFS interaction.

\begin{figure}[h]
\centering
\includegraphics*[scale=0.35]{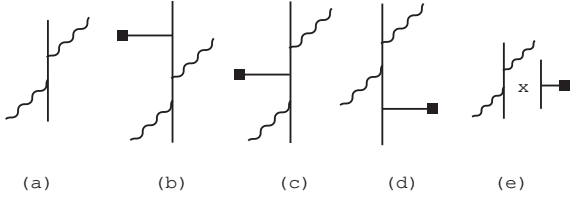}
\caption{Contributions to the dynamic polarizability
$\alpha(\omega)$. Diagram (a) represents the second-order
contribution arising from two photon interactions (wavy lines).
Diagrams (b-e) represent the additional effect of the hyperfine
interaction (capped line) and correspond respectively to the
third-order top, center, bottom, and normalization terms as
described in the text.} \label{diagrams}
\end{figure}

Formally, this effect appears in the third-order double perturbation
theory with two laser and one HFS interactions. We build the
perturbation theory in terms of combined interaction
\[
V=V_{\mathrm{hfs}}+V_{E1}\left(  t\right)  .
\]
The convenience of the Floquet formalism is that we may immediately
employ the conventional formula for the third-order energy
correction
\begin{eqnarray*}
\delta E_{a}^{\left[  2+1\right]  }&=&\sum_{b,c\neq a}\frac{V_{ab}V_{bc}V_{ca}%
}{\left(  E_{b}^{\left(  0\right)  }-E_{a}^{\left(  0\right)  }\right)
\left(  E_{c}^{\left(  0\right)  }-E_{a}^{\left(  0\right)  }\right)  }%
\\&&-V_{aa}\sum_{b\neq a}\frac{V_{ab}V_{ba}}{\left(  E_{b}^{\left(
0\right)
}-E_{a}^{\left(  0\right)  }\right)  ^{2}},%
\end{eqnarray*}
Here  $a,b,c$ are dressed atomic states, i.e.
$|a\rangle=|\psi_{a}\rangle e^{in\omega t}$, with $n$ representing
the number of photons ($n$ could be both negative and positive). The
scalar product, in addition to the conventional Hilbert space
operational definition includes an averaging over the period of
oscillation of the laser field. Explicitly, after the time averaging
(now $a,b,$ and $c$ are atomic states)
\[
\delta E_{a}^{\left[  2+1\right]  }\left(  \omega\right)
=T_{a}\left( \omega\right)  +C_{a}\left(  \omega\right)
+B_{a}\left(  \omega\right) +O_{a}\left(  \omega\right),
\]%
\begin{eqnarray*}
T_{a}\left(  \omega\right)   &  =&\langle
a|V_{\mathrm{hfs}}R_{E_{a}}\left( 0\right)  vR_{E_{a}}\left(
\omega\right)  v^{\dagger}|a\rangle\\&&+\langle
a|V_{\mathrm{hfs}}R_{E_{a}}\left(  0\right)
v^{\dagger}R_{E_{a}}\left(
-\omega\right)  v|a\rangle,\\
C_{a}\left(  \omega\right)   &  =&\langle a|vR_{E_{a}}\left(
\omega\right) V_{\mathrm{hfs}}R_{E_{a}}\left(  \omega\right)
v^{\dagger}|a\rangle\\&&+\langle a|v^{\dagger}R_{E_{a}}\left(
-\omega\right)  V_{\mathrm{hfs}}R_{E_{a}}\left(
-\omega\right)  v|a\rangle,\\
B_{a}\left(  \omega\right)   &  =&\left[  T_{a}\left(  \omega\right)
\right]
^{\ast},\\
O_{a}\left(  \omega\right)   &  =&-\left(  V_{\mathrm{hfs}}\right)
_{aa}\left(  \langle a|v^{\dagger}\left(  R_{E_{a}}\left(
\omega\right) \right)  ^{2}v|a\rangle\right.\\&&\left.+\langle
a|v\left( R_{E_{a}}\left(  -\omega\right) \right)
^{2}v^{\dagger}|a\rangle\right)  .
\end{eqnarray*}
Here $T_{a}\left(  \omega\right)  $, $C_{a}\left(  \omega\right)  $,
and $B_{a}\left(  \omega\right)  $ stand for top, center, and bottom
position of the HFS interaction in the respective diagram. The term
$O_{a}\left( \omega\right)  $ describes the normalization term. The
relevant diagrams are shown in Fig.~\ref{diagrams}.

The interaction of the electron with the nuclear magnetic moment $\mu$ reads%
\[
V_{\mathrm{hfs}}=\left(  \mu\cdot\mathcal{T}^{\left(  1\right)  }\right)  ,
\]
where $\mathcal{T}^{\left(  1\right)  }$ is the rank 1 irreducible tensor
operator acting in the electronic coordinates with components
\[
\mathcal{T}_{\lambda}^{\left(  1\right)  }=-\frac{\left\vert e\right\vert
}{4\pi\varepsilon_{0}}\frac{i\sqrt{2}\left(  \mathbf{\alpha}\cdot
\mathbf{C}_{1\lambda}^{\left(  0\right)  }\left(  \mathbf{\hat{r}}\right)
\right)  }{cr^{2}},
\]
where $\mathbf{\alpha}$ stands for the  Dirac matrices and
$\mathbf{C}_{1\lambda}^{\left(  0\right)  }\left(
\mathbf{\hat{r}}\right)  $ are the normalized vector spherical
harmonics. In the formulas below we require the reduced matrix
element of the nuclear moment operator in the nuclear basis,
$\langle I||\mu||I\rangle$. It is related to the nuclear magnetic
$g$-factor as
\[
\langle I||\mu||I\rangle=\frac{1}{2}\sqrt{\left(  2I\right)  \left(
2I+1\right)  \left(  2I+2\right)  }~g\mu_{n},
\]
 $\mu_{n}$ being the nuclear magneton.

Carrying out the angular decomposition similar to the second-order
analysis of the preceding section we find that  the expressions for
the shift, Eq.~(\ref{Eq:StarkEnergyShift}), remain the same with the
reduced polarizabilities $\alpha_{nF}^{\left(  K\right)  }\left(
\omega\right)  $ redefined as
\[
\alpha_{nF}^{\left(  K\right)  }\left(  \omega\right)  \rightarrow\alpha
_{nF}^{\left(  K\right)  }\left(  \omega\right)  +\beta_{nF}^{\left(
K\right)  }\left(  \omega\right)  .
\]
The third-order rank $K=0,1,2$ corrections are given by the sum over
contributions from diagrams of Fig.~\ref{diagrams}.
\[
\beta_{nF}^{\left(  K\right)  }\left(  \omega\right)  =2\beta_{nF}^{\left(
K\right)  }\left(  \omega;\mathrm{T}\right)  +\beta_{nF}^{\left(  K\right)
}\left(  \omega;\mathrm{C}\right)  +\beta_{nF}^{\left(  K\right)  }\left(
\omega;\mathrm{O}\right)  .
\]
Explicitly,
\begin{widetext}
\begin{align*}
\beta_{nF}^{\left(  K\right)  }\left(  \omega;\mathrm{T}\right)
=[F]\sqrt{[K]}
\sum_{J_{a}J_{b}}\left(  -1\right)  ^{J+J_{a}}\left\{
\begin{tabular}
[c]{lll}%
$I$ & $I$ & $1$\\
$J_{a}$ & $J$ & $F$%
\end{tabular}
\ \right\}  \left\{
\begin{tabular}
[c]{lll}%
$J$ & $1$ & $J_{b}$\\
$1$ & $J_{a}$ & $K$%
\end{tabular}
\ \right\}  \left\{
\begin{tabular}
[c]{lll}%
$K$ & $J$ & $J_{a}$\\
$I$ & $F$ & $F$%
\end{tabular}
\ \ \right\}  T_{J_{a}J_{b}}^{\left(  K\right)  }\left(  nJ,\omega\right)  ,
\end{align*}%
\begin{align*}
\beta_{nF}^{\left(  K\right)  }\left(  \omega;\mathrm{C}\right)
=[F]\sqrt{[K]} &  \sum_{J_{a}J_{b}}\sum_{J_{i}}[J_{i}](-1)^{2J_{a}+J_{b}%
+J}
\left\{
\begin{array}
[c]{ccc}%
J & J & J_{i}\\
I & I & 1\\
F & F & K
\end{array}
\right\}  \left\{
\begin{array}
[c]{ccc}%
J & J & J_{i}\\
J_{a} & J_{b} & 1\\
1 & 1 & K
\end{array}
\right\}  C_{J_{a}J_{b}}^{\left(  K\right)  }\left(  nJ,\omega\right)  ,
\end{align*}%
\begin{align*}
\beta_{nF}^{\left(  K\right)  }\left(  \omega;\mathrm{O}\right)   &
=\left( -1\right)  ^{2J+1}\left[  F\right]  \sqrt{\left[  K\right]
}\left\{
\begin{array}
[c]{ccc}%
1 & I & I\\
F & J & J
\end{array}
\right\}
\left\{
\begin{array}
[c]{ccc}%
J & J & K\\
F & F & I
\end{array}
\right\}  \sum_{J_{a}}\left\{
\begin{array}
[c]{ccc}%
K & J & J\\
J_{a} & 1 & 1
\end{array}
\right\}
O_{J_{a}}^{\left(  K\right)  }\left(  nJ,\omega\right)  .
\end{align*}
Here we introduced the reduced sums
\begin{eqnarray}
T_{J_{a}J_{b}}^{\left(  K\right)  }\left(  nJ,\omega\right)
&=&\langle
I||\mu||I\rangle\sum_{n_{b}}\sum_{n_{a}\neq n}   \langle nJ||\mathcal{T}%
^{\left(  1\right)  }||n_{a}J_{a}\rangle\langle n_{a}J_{a}||D||n_{b}%
J_{b}\rangle\langle n_{b}J_{b}||D||nJ\rangle \nonumber \\
&& \times \left(  \frac{1}{E-E_{a}}\frac{1}{E-E_{b}+\omega}+\left(
-1\right) ^{K}\left(  \omega\rightarrow-\omega\right)  \right),
\label{Eq:TopSum}
\end{eqnarray}%
\begin{eqnarray}
C_{J_{a}J_{b}}^{\left(  K\right)  }\left(  nJ,\omega\right)
&=&\langle I||\mu||I\rangle\sum_{n_{a}n_{b}}\langle
nJ||D||n_{a}J_{a}\rangle   \langle n_{a}J_{a}||\mathcal{T}^{\left(
1\right)  }||n_{b}J_{b}\rangle\langle n_{b}J_{b}||D||nJ\rangle
\nonumber \\&&\times \left(
\frac{1}{E-E_{a}+\omega}\frac{1}{E-E_{b}+\omega}+\left( -1\right)
^{K}\left(  \omega\rightarrow-\omega\right)  \right),
\label{Eq:CenterSum}
\end{eqnarray}%
\begin{eqnarray}
O_{J_{a}}^{\left(  K\right)  }\left(  nJ,\omega\right) &=&\langle
I||\mu||I\rangle\langle nJ||\mathcal{T}^{\left(  1\right)
}||nJ\rangle\sum_{n_{a}}\langle nJ||D||n_{a}J_{a}\rangle\langle
n_{a}J_{a}||D||nJ\rangle
\left( \frac {1}{\left(
E-E_{a}+\omega\right)  ^{2}}+\left( -1\right) ^{K}\left(
\omega\rightarrow-\omega\right)  \right). \label{Eq:NormSum}
\end{eqnarray}
\end{widetext}
Notice that the angular momenta of the intermediate states $J_{a}$
and $J_{b}$ are fixed.

\subsection{Numerical evaluation}
\label{calculations}
To perform the calculations we use an {\em ab initio} approach
which has been described in detail in Ref.~\cite{AngDzuFla06}.
In this approach high accuracy is attained by including
important many-body and relativistic effects.

Calculations start from the relativistic Hartree-Fock (RHF) method
in the $V^{N-1}$ approximation. This means that the initial RHF
procedure is done for a closed-shell atomic core with the valence
electron removed. After that, the states of the external electron
are calculated in the field of the frozen core. Correlations are
included by means of the correlation potential
method~\cite{DzuFlaSil87}. We use the all-order correlation
potential $\hat \Sigma$ which includes two classes of the
higher-order terms: screening of the Coulomb interaction and
hole-particle interaction (see, e.g.,~\cite{DzuFlaSus89} for
details).

To calculate $\hat \Sigma$ we need a complete set of single-electron
orbitals. We use the B-spline technique~\cite{JohBluSap88} to
construct the basis. The orbitals are built as linear combinations of
40 B-splines of order 9 in a cavity of radius 40$a_B$.
The coefficients are chosen from the condition that the
orbitals are the eigenstates of the RHF Hamiltonian $\hat H_0$ of the
closed-shell core. The $\hat \Sigma$ operator is
calculated with the technique which combines solving equations for
the Green functions (for the direct diagram) with the summation over
complete set of states (exchange diagram)~\cite{DzuFlaSus89}.

The correlation potential $\hat \Sigma$ is then used to build a new
set of single-electron states, the so-called Brueckner orbitals.
This set is to be used in the summation in equations
(\ref{Eq:TopSum}), (\ref{Eq:CenterSum}), and (\ref{Eq:NormSum}).
Here again we use the B-spline technique to build the basis. The
procedure is very similar to the construction of the RHF B-spline
basis. The only difference is that the new orbitals are now the
eigenstates of the $\hat H_0 + \hat \Sigma$ Hamiltonian.

Brueckner orbitals which correspond to the lowest valence states are
good approximations to the real physical states. Their quality can
be tested by comparing experimental and theoretical energies.
Moreover, their quality can be further improved by rescaling the
correlation potential $\hat \Sigma$ to fit the experimental energies
exactly. We do this by replacing the $\hat H_0 + \hat \Sigma$
Hamiltonian with $\hat H_0 + \lambda \hat \Sigma$, in which the
rescaling parameter $\lambda$ is chosen for each partial wave to fit
the energy of the first valence state. The values of $\lambda$ are
$\lambda_s=1$, $\lambda_p=0.97$, and $\lambda_d=0.95$. Note that
these values are very close to unity. This means that even without
rescaling the accuracy is good and only a small adjustment of $\hat
\Sigma$ is needed. Note also that since the rescaling procedure
affects not only energies but also the wave functions, it usually
leads to improved values of the matrix elements of external fields.
In fact, this is a semi-empirical method to include omitted
higher-order correlation corrections.

Matrix elements of the HFS and electric dipole operators are found
by means of the time-dependent Hartree-Fock (TDHF)
method~\cite{DzuFlaSil87,DzuFlaSus84}. This method is equivalent to
the well-known random-phase approximation (RPA). In the TDHF method,
the single-electron wave functions are presented in the form $\psi =
\psi_0 + \delta \psi$, where $\psi_0$ is the unperturbed wave
function. It is an eigenstate of the RHF Hamiltonian $\hat H_0$:
$(\hat H_0 -\epsilon_0)\psi_0 = 0$.  $\delta \psi$ is the correction
due to an external field. It can be found by solving the TDHF
equation
\begin{equation}
    (\hat H_0 -\epsilon_0)\delta \psi = -\delta\epsilon \psi_0 - \hat F \psi_0 -
  \delta \hat V^{N-1} \psi_0,
  \label{TDHF}
\end{equation}
where $\delta\epsilon$ is the correction to the energy due to the
external field ($\delta\epsilon\equiv 0$ for the electric dipole
operator), $\hat F$ is the operator of the external field
($V_\mathrm{hfs}$ or $-\mathbf{D}\cdot \mathcal{E}$), and $\delta
\hat V^{N-1}$ is the correction to the self-consistent potential of
the core due to the external field.

The TDHF equations are solved self-consistently for all states in the core. Then the
matrix elements between any (core or valence) states $n$ and $m$ are given by
\begin{equation}
    \langle \psi_n | \hat F + \delta \hat V^{N-1} | \psi_m \rangle.
    \label{mel}
\end{equation}
The best results are achieved when $\psi_n$ and $\psi_m$ are the
Brueckner orbitals computed with rescaled correlation potential
$\lambda\hat{\Sigma}$.

We use equation (\ref{mel}) for all HFS and electric dipole matrix
elements in evaluating the top, center, bottom, and normalization
diagrams
(Eqs.~(\ref{Eq:TopSum}),(\ref{Eq:CenterSum}),(\ref{Eq:NormSum})),
except for the ground state HFS matrix element in the normalization
diagram where we use experimental data. The results are presented in
section \ref{results}.

\section{Experiment}
We measure the frequency shift of the Cs clock transition ($| F=3,
M_F=0 \rangle $ to $|{ F}=4, { M_F}=0 \rangle$) induced by a far
detuned laser beam. A Cs fountain clock \cite{Bize} is used. At each
cycle $\sim10^6$ atoms are loaded in an optical molasses and cooled
to below $2\,\mu$K. Moving molasses launches the atoms upwards with
a speed of $4.1$~m/s where they pass twice through a microwave
cavity thereby realizing Ramsey spectroscopy. The Zeeman degeneracy
is lifted by a 1.6~mG magnetic field aligned along the fountains
axis. The detuned laser beam is a traveling wave beam also aligned
on the axis of the fountain. The light polarization is linear with
respect to the light propagation. The beam waist is larger than the
11~mm diameter opening in the microwave cavity. This assures that
all atoms passing through the opening and being detected experience
the light. The light intensity averaged over 1 cm$^2$ is measured by
a commercial powermeter (Newport 840-C) before entering the
fountains vacuum chamber. One intensity measurement is taken before
and one after each one-day run. The intensity drift between the two
is of the order of 1\%, however the error of the light intensity
experienced by each atom is rather high. This is due to our
ignorance of the exact intensity distribution, the exact atom
distribution and intensity losses in the vacuum window as well as
parasite reflections inside the vacuum chamber. We estimate the
intensity error as 20\%.

The frequency shift is measured by alternating the fountain's
configuration every 50 cycles. The first configuration is the
standard clock operation. The second configuration is identical to
the first plus the laser beam opened during the Ramsey period. This
assures that the atom preparation and cooling is not disturbed by
the  laser and that the atomic cloud is identical in the two
configurations. Hence, we can assume that all other clock shifts, in
particular collisions, are identical for the two configurations. The
absolute frequency is measured for each configuration against a
highly stable local oscillator exhibiting no significant drift
during several hundred cycles. The frequency shift induced by the
light calculates as the simple difference
\[
\delta \nu = \nu_{\rm with~light} - \nu_{\rm without~light}.
\]
The frequency shift is measured for a number of laser intensities.
Two sets of measurements are taken for light at wavelengths of 532
nm and 780 nm. The averages of each set weighted by the statistical
frequency uncertainty give a light shift of $(-3.51 \pm 0.7)\times
10^{-4}$ Hz(W/cm$^2$)$^{-1}$ for 532 nm and $(-2.27 \pm 0.4)\times
10^{-2}$ Hz(W/cm$^2$)$^{-1}$ for 780 nm. The statistical uncertainty
is negligible before the uncertainty on the light intensity.

\section{Results and discussion}

\label{results}

The shift of the clock frequency is given by (cf. Eq.~(\ref{Eq:ClockShift}))
\[
\delta \nu_L^\mathrm{Stark}=-\left(  \frac{1}{2}\mathcal{E}_L\right)  ^{2}\ \delta\alpha\left(
\omega_L\right),
\]
where $\delta\alpha\left(\omega_L\right)$ is the differential polarizability.
The conversion factor between differential polarizability in atomic units and the ratio of the shift to
laser intensity in practical units is given by
\[
\frac{\delta \nu_L^\mathrm{Stark}\left[  \mathrm{Hz}\right]  }{I_L\left[  \frac{\mathrm{mW}%
}{\mathrm{cm}^{2}}\right]  }=-4.68\times10^{-5}~\times~\delta\alpha\left(
\omega\right)  \left[  \mathrm{a.u.}\right] \, .
\]
We start by considering DC polarizabilities. In the static regime
($\omega_L = 0$), our calculations give $\delta \alpha = 1.82 \times
10^{-2}\,\mathrm{a.u.}$, which translates into the commonly used DC
Stark coefficient $k_S =-2.26 \times 10^{-10} \,
\mathrm{Hz}/(\mathrm{V/m})^2$. Notice that this value includes only
the scalar part of the polarizability. This is in agreement with the
most accurate experimental result~\cite{SimLauCla98} of  $k_S
=-2.271(4) \times 10^{-10} \, \mathrm{Hz}/(\mathrm{V/m})^2$.

\begin{figure}[ht]
\begin{center}
\includegraphics*[scale=0.4]{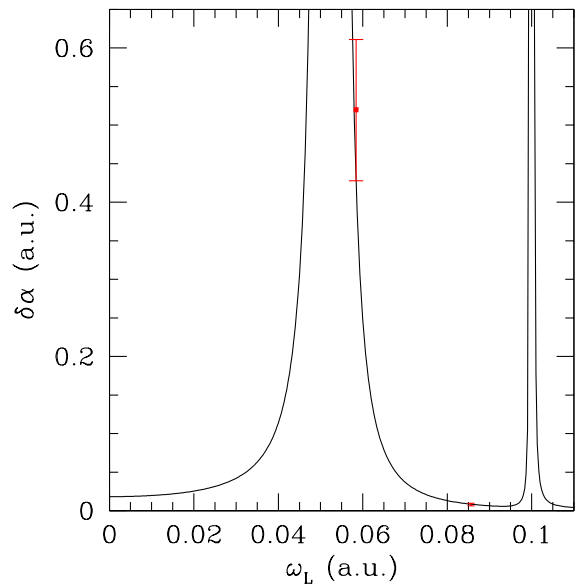}
\end{center}
\caption{(Color online) Differential polarizability of the Cs clock
transition ($M_F=M_F'=0$) in the
$\mathbf{B}\parallel\mathbf{\hat{k}}$ configuration as a function of
the probe laser frequency. Two experimental measurements (at 780 nm
and 532 nm) are compared with theoretical predictions (solid curve).
Refer to Fig.~\ref{Fig:ClockShift2} for better graphical resolution
of the 532 nm experimental point.} \label{Fig:ClockShift}
\end{figure}

\begin{figure}[ht]
\begin{center}
\includegraphics*[scale=0.4]{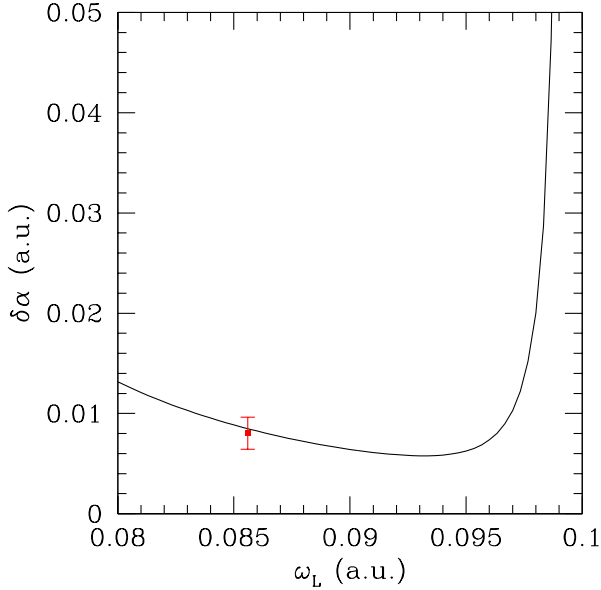}
\end{center}
\caption{(Color online) Same as in Fig.~\ref{Fig:ClockShift}, in the
region of the experimental point at 532 nm. }
\label{Fig:ClockShift2}
\end{figure}

For the AC case, our calculated differential polarizability $\delta
\alpha(\omega_L)$ for the cesium clock transition is presented in
Fig.~\ref{Fig:ClockShift} as a function of laser frequency. Both
values are given in atomic units. The two peaks correspond to the
$6s - 6p$ and $6s - 7p$ resonances. The graph never crosses zero,
which implies no magic frequency. Experimental results for two laser
wavelengths are also shown (also see Fig.~\ref{Fig:ClockShift2}).
Calculated and experimental relative frequency shifts are compared
in Table~\ref{tab:ex} and found to be in agreement with each other.


\begin{table}[ht]
\caption{Comparison of the theoretical and experimental AC frequency
shifts for the clock transition in Cs} \label{tab:ex}
\begin{ruledtabular}
\begin{tabular}{cccc}
 $\lambda$,[nm] & $\omega$,[a.u.] & \multicolumn{2}{c}{$\delta \nu_L/I_L$, [Hz/mW/cm$^2$]} \\
                &                 & Theor.  & Expt. \\
\hline
780 & 0.0584 & $-1.95 \times 10^{-2}$ & $-2.27(40) \times 10^{-2}$ \\
532 & 0.0856 & $-3.73 \times 10^{-4}$ & $-3.51(70) \times 10^{-4}$ \\
\end{tabular}
\end{ruledtabular}
\end{table}

To summarize, we presented a comprehensive analysis of the AC Stark
shift of the Cs microwave atomic clock transition. Theoretical
analysis based on the second and third order perturbation theory is
accompanied by measurements. Calculations and measurements are in
good agreement with each other and indicate the absence of a magic
frequency at least for the $M_F=0$ clock levels with zero
projections of the total angular momentum on the quantizing magnetic
field.

\appendix*
\section{}
Considering the complexities of working with the angular algebra,
here we analyze Eq.~(2) of Ref.~\cite{ZhoCheChe05} for the 2nd-order
dynamic Stark shift. Starting from their equation we again show that
there is no AC Stark shift of the clock transition frequency.
Eq.~(2) of Ref.~\cite{ZhoCheChe05} for the polarizability contains a
summation over $M^\prime$, which we manipulate
\begin{widetext}
\begin{eqnarray} \sum_{M^\prime}\left(
\begin{array}{ccc}
 F^\prime & 1 &  F \\
 M^\prime & p & -M
\end{array}
\right)^2&=&\left(-1\right)^{F^\prime-M+p}\sum_{M^\prime}\left(
\begin{array}{ccc}
 F &  1 &  F^\prime \\
 M & -p & -M^\prime
\end{array}
\right)\left(
\begin{array}{ccc}
 F^\prime & 1 &  F \\
 M^\prime & p & -M
\end{array}
\right)\nonumber\\
&=&\left(-1\right)^{F^\prime-M+p}\left(-1\right)^{2F}\sum_{KQ}
\left(-1\right)^{K-Q}\left[K\right]\left(
\begin{array}{ccc}
 F &  F &  K \\
 M & -M & -Q
\end{array}
\right)\left(
\begin{array}{ccc}
 K & 1 &  1 \\
 Q & p & -p
\end{array}
\right)\left\{
\begin{array}{ccc}
 1 & 1 & K \\
 F & F & F^\prime
\end{array}
\right\}\label{Eq:sumrule}\qquad\\
&=&\left(-1\right)^{F-M+p}\sum_{K}
\left(-1\right)^{F+F^\prime+K}\left[K\right]\left(
\begin{array}{ccc}
  F & K & F \\
 -M & 0 & M
\end{array}
\right)\left(
\begin{array}{ccc}
  1 & K & 1 \\
 -p & 0 & p
\end{array}
\right)\left\{
\begin{array}{ccc}
 1 & 1 & K \\
 F & F & F^\prime
\end{array}
\right\}\label{Eq:Q0}.
\end{eqnarray}
The expression~(\ref{Eq:sumrule}) is obtained from using the
summation rule 12.1(5) of Ref.~\cite{VarMosKhe88}; we further obtain
expression~(\ref{Eq:Q0}) by noting that only $Q=0$ terms are
non-zero in the summation.

Now we take the $F^\prime$-dependent part of~(\ref{Eq:Q0}) with the
$F^\prime$-dependent part of Eq.~(2) of Ref.~\cite{ZhoCheChe05} and
take the summation over $F^\prime$
\begin{eqnarray}
\sum_{F^\prime}\left(-1\right)^{F+F^\prime+K}\left[F^\prime\right]\left\{
\begin{array}{ccc}
 J        & J^\prime & 1 \\
 F^\prime & F        & I
\end{array}
\right\}^2\left\{
\begin{array}{ccc}
 1 & 1 & K \\
 F & F & F^\prime
\end{array}
\right\}&=&\sum_{F^\prime}\left(-1\right)^{F+F^\prime+K}\left[F^\prime\right]\left\{
\begin{array}{ccc}
 1 & F        & F^\prime \\
 I & J^\prime & J
\end{array}
\right\}\left\{
\begin{array}{ccc}
 I & J^\prime & F^\prime \\
 1 & F        & J
\end{array}
\right\}\left\{
\begin{array}{ccc}
 1 & F & F^\prime \\
 F & 1 & K
\end{array}
\right\}\nonumber\\
&=&\left(-1\right)^{I-J^\prime+F}\left\{
\begin{array}{ccc}
 J & J & K \\
 1 & 1 & J^\prime
\end{array}
\right\}\left\{
\begin{array}{ccc}
 J & J & K \\
 F & F & I
\end{array}
\right\}\label{Eq:sumrule2}.
\end{eqnarray}
The expression~(\ref{Eq:sumrule2}) is obtained from using the
summation rule 9.8(6) of Ref.~\cite{VarMosKhe88}.

Combining the above results gives
\begin{eqnarray}
\sum_{F^\prime M^\prime} \left[F\right]\left[F^\prime\right] \left(
\begin{array}{ccc}
 F^\prime & 1 &  F \\
 M^\prime & p & -M
\end{array}
\right)^2\left\{
\begin{array}{ccc}
 J        & J^\prime & 1 \\
 F^\prime & F        & I
\end{array}
\right\}^2&=&\left(-1\right)^{F-M+p}\left(-1\right)^{I-J^\prime+F}
\left[F\right]\sum_{K}\left[K\right]\left(
\begin{array}{ccc}
  F & K & F \\
 -M & 0 & M
\end{array}
\right)\left(
\begin{array}{ccc}
  1 & K & 1 \\
 -p & 0 & p
\end{array}
\right)\nonumber\\
&&\times\left\{
\begin{array}{ccc}
 J & J & K \\
 1 & 1 & J^\prime
\end{array}
\right\}\left\{
\begin{array}{ccc}
 J & J & K \\
 F & F & I
\end{array}
\right\}.\nonumber
\end{eqnarray}
\end{widetext}
Not surprisingly, the six-$j$ symbols here are identical to the ones
appearing in the previously derived polarizabilities
$\alpha^{(K)}_{nF}(\omega)$. Hence, this makes the connection to
scalar ($K=0$), vector ($K=1$), and tensor ($K=2$) parts.

First we focus on the case $p=0$; this corresponds to linear
polarization in the $\mathbf B
\parallel \hat \varepsilon$ geometry. For $J=1/2$ atomic states the
tensor part ($K=2$) is necessarily zero due to selection rules in
the six-$j$ symbols (this is the case regardless of polarization).
Furthermore the vector part ($K=1$) is zero due to the fact that the
top row of the second three-$j$ symbol sums to an odd number (or see
(\ref{Eq:3j}) below, with $p=0$). This leaves us with only the
scalar part ($K=0$) to analyze. In this case, the r.h.s.~simply
reduces to $1/(3[J])=1/6$. Thus we can conclude that for linear
polarization of this geometry, the 2nd-order dynamic Stark shift is
$F$-independent for $J=1/2$ atomic states.

For the $\mathbf B
\parallel \mathbf{\hat{k}}$ geometry, the linear polarization is regarded as
an equal mixture of $\sigma^+$ ($p=+1$) and $\sigma^-$ ($p=-1$)
circularly polarized light. Again the tensor part is necessarily
zero for $J=1/2$. For the vector part, we note the three-$j$ symbol
relation
\begin{equation}
\left(
\begin{array}{ccc}
  1 & K & 1 \\
 -p & 0 & p
\end{array}
\right)=\left(-1\right)^K\left(
\begin{array}{ccc}
  1 & K &  1 \\
  p & 0 & -p
\end{array}
\right).\label{Eq:3j}
\end{equation}
Thus, when we take equal mixtures of $\sigma^+$ and $\sigma^-$
light, the vector contribution drops out. Again, we are left with
only the scalar part. Not surprisingly, we again obtain the result
$1/(3[J])=1/6$ when taking equal mixtures of $\sigma^+$ and
$\sigma^-$ light.

The above results then generalize to any geometry for linearly
polarized light.

\begin{acknowledgments}

This work was supported in part by the US National Science
Foundation, by the Australian Research Council and by the US
National Aeronautics and Space Administration under
Grant/Cooperative Agreement No. NNX07AT65A issued by the Nevada NASA
EPSCoR program. The SYRTE is Unit\'e de Recherche of the
Observatoire de Paris and the Universit\'e Pierre et Marie Curie
associated to the CNRS (UMR8630). It is a national metrology
laboratory of the Laboratoire National de M\'etrologie et d'Essais
(LNE) and member of the Institut francilien de recherche sur les
atomes froids (IFRAF). S.G. acknowledges travel support from the
LNE.
\end{acknowledgments}



\end{document}